\RequirePackage{booktabs}

\documentclass[referee,sn-basic]{sn-jnl}
\usepackage{graphicx}%
\usepackage{multirow}%
\usepackage{amsmath,amssymb,amsfonts}%
\usepackage{amsthm}%
\usepackage{mathrsfs}%
\usepackage[title]{appendix}%
\usepackage{xcolor}%
\usepackage{textcomp}%
\usepackage{manyfoot}%
\usepackage{booktabs}%
\usepackage{algorithm}%
\usepackage{algorithmicx}%
\usepackage{algpseudocode}%
\usepackage{listings}%
\usepackage{algorithm}
\usepackage{algpseudocode}
\usepackage{epstopdf}
\usepackage{soul}
\usepackage[style=authoryear-ibid,backend=biber]{biblatex}
\usepackage{caption}
\usepackage{epstopdf}
\usepackage{lipsum}
\usepackage{soul}
\usepackage{bm}
\usepackage{bbm}
\usepackage{placeins}
\usepackage[english]{babel}
\usepackage{csquotes}
\usepackage[usestackEOL]{stackengine} 
\usepackage{multirow}
\usepackage{tabulary}
\usepackage{amsthm}
\usepackage{graphicx}
\usepackage{booktabs}
\usepackage{dcolumn}

\theoremstyle{plain} 
\newtheorem{theorem}{Theorem}
\newtheorem{proposition}[theorem]{Proposition}

\usepackage{arydshln}
\theoremstyle{thmstylethree}%
\newtheorem{definition}{Definition}%

\begin{document}

\title[The Population Resemblance Statistic]{\vspace{-0.5in}\linespread{0.9}\selectfont The Population Resemblance Statistic: A Chi-Square Measure of Fit for Banking}

\author[1,4]{\fnm{C.J.} \sur{Potgieter}}\email{c.potgieter@tcu.edu}
\equalcont{These authors contributed equally to this work.}

\author[2,3,5]{\fnm{C.} \sur{Van Zyl}}\email{corli.vanzyl@nwu.ac.za}
\equalcont{These authors contributed equally to this work.}

\author*[2,3]{\fnm{W.D.} \sur{Schutte}}\email{wd.schutte@nwu.ac.za}
\equalcont{These authors contributed equally to this work.}

\author[4,6]{\fnm{F.} \sur{Lombard}} \nomail
\equalcont{These authors contributed equally to this work.}

\affil[1]{\small\orgdiv{Department of Mathematics}, \orgname{Texas Christian University}, \orgaddress{\city{Fort Worth}, \state{Texas}, \country{USA}}}

\affil*[2]{\small\orgdiv{Centre for Business Mathematics and Informatics}, \orgname{North-West University},  \orgaddress{\city{Potchefstroom}, \country{South Africa}}}

\affil[3]{\small\orgname{Absa Bank Limited},  \orgaddress{\city{Sandton}, \country{South Africa}}}

\affil[4]{\small\orgdiv{Department of Statistics}, \orgname{University of Johannesburg}, \orgaddress{\city{Johannesburg}, \country{South Africa}}}

\affil[5]{\small\orgname{Afrimat Limited}, \orgaddress{\city{Cape Town}, \country{South Africa}}}

\affil[6]{\small\orgdiv{Posthumous}}

\abstract{The Population Stability Index (PSI) is a widely used measure in credit risk modeling and monitoring within the banking industry. Its purpose is to monitor for changes in the population underlying a model, such as a scorecard, to ensure that the current population closely resembles the one used during model development. If substantial differences between populations are detected, model reconstruction may be necessary. Despite its widespread use, the origins and properties of the PSI are not well documented. Previous literature has suggested using arbitrary constants as a rule-of-thumb to assess resemblance (or ``stability''), regardless of sample size. However, this approach too often calls for model reconstruction in small sample sizes while not detecting the need often enough in large sample sizes.

This paper introduces an alternative discrepancy measure, the Population Resemblance statistic (PRS), based on the Pearson chi-square statistic. Properties of the PRS follow from the non-central chi-square distribution. Specifically, the PRS allows for critical values that are configured according to sample size and the number of risk categories. Implementation relies on the specification of a set of parameters, enabling practitioners to calibrate the procedure with their risk tolerance and sensitivity to population shifts. The PRS is demonstrated to be universally competent in a simulation study and with real-world examples.
}


\maketitle

\section{Introduction}
\label{sec_intro}

Testing the stability of a population used for model development is common practice in model risk management. In credit risk modeling, the Population Stability Index (PSI) is the most widely used measure to monitor the evolution of the population underlying a model, through assessing the degree of discrepancy, conversely similarity, between two discrete probability distributions, see \textcite[pp. 155 ff.]{thomas2002credit} and \textcite[pp. 368 ff.]{siddiqi2017intelligent}. Small deviations in the population can result in inaccurate or unreliable model predictions. {For instance, when modeling the probability of default (or related quantities such as risk score, exposure at default, or loss given default) based on a given population of borrowers, the reliability of the model becomes questionable when the population changes substantively.} 

{Unsurprisingly, many prudential authorities require that the current population \emph{resemble} (emphasis ours) the one used during regulatory capital model development} (see \textcite{ecb2024internalmodels, ecb2019validation}, \textcite{federalreserve2023sr1107a1}, \textcite{resbank2022credit}; and \textcite{pruitt2010applied} for an application of the PSI in SAS). Other areas of application include insurance, healthcare, engineering, and marketing (see \textcite{huang}, \textcite{li}, \textcite{sahu}, \textcite{dong2022prediction}, \textcite{wu2010enterprise}, \textcite{mcadams2022risk}, \textcite{chou2022expert}, \textcite{karakoulas2004empirical} and \textcite{brockett1995information}).

Despite its widespread use, the origins and properties of the PSI are not widely understood. The PSI is based on the Kullback-Leibler divergence, measuring difference between two probability distributions \parencite[Eq.~2.6]{kullback1951information}. The earliest reference to {the PSI} measure can be found in \textcite{lewis1994introduction}, who also coined the term ``Population Stability Index'' and {popularized use of} {the so-called} \textit{Lewis constants} as thresholds (see \textcite[p.~155 ff.]{thomas2002credit}, \textcite[p.~368 ff.]{siddiqi2017intelligent}). \textcite[p.~106]{lewis1994introduction} describes the PSI without formulating a hypothesis in the statistical sense but notes, ``If a user finds the distribution of scores \textit{close together}, [they] can be confident that the population has not changed.'' In his example, a PSI below 0.10 indicates that the current population is sufficiently similar to the original and no action is required, a value between 0.10 and 0.25 suggests that call for further investigation, and a value above 0.25 signals a material change in the incoming population that may necessitate model reconstruction. {The specification of these constants neglects any connection to a defined magnitude of population shift or to the role of sampling variability. By contrast, our framework introduces a precise definition of distributional similarity, which we refer to as \textit{resemblance}, and explicitly links this to statistical thresholds that specify an associated maximum tolerable population shift.} 

The arbitrary nature of the Lewis constants has been acknowledged by authors and practitioners alike; see \textcite{yurdakul2020statistical},  \textcite{dupisanievisagie2020}, and \textcite{moodys2021email}. These thresholds pose limitations in small portfolios, where the shift of a single borrower results in a distortion of the PSI beyond its thresholds and unnecessarily prompts model reconstruction\footnote{See \textcite[p.~71]{nedbankp3}, \textcite[p.~51]{stbankp3} and \textcite[p.~232]{frp3} for the prevalence of portfolios with less than (e.g.) 100 borrowers.}. Conversely, in large portfolios (e.g., retail books with over one million borrowers), substantial shifts may go undetected. Together, these examples highlight the PSI’s limitations as a one-size-fits-all discrepancy measure and highlight the need for more adaptable and interpretable approaches.

\subsection{{Contributions of this paper}}\label{sec:contrib}

{This paper introduces the Population Resemblance Statistic (PRS), a new procedure for monitoring changes in categorical population proportions.  Rooted in the Pearson chi-square framework and drawing on properties of the non-central chi-square distribution, the PRS is centered on the concept of $\delta$-resemblance --- a formalization of tolerable population shift that allows practitioners to tune the procedure to a specified level of risk tolerance. The resulting critical values for decision-making are directly determined by the sample size and the number of risk categories, making the approach adaptable across a range of settings.}

Importantly, the PRS is not merely a marginal refinement of existing tools but represents a conceptual shift in the monitoring paradigm. We move away from the traditional approach, which applies two significance levels (or, in some cases, arbitrary thresholds) to a test statistic under a single point null hypothesis assuming exact population equality. {This traditional formulation leads to two well-known problems: in large samples, even negligible shifts are flagged as significant; in small samples, substantial changes may go undetected due to low statistical power -- see \textcite{schuirmann1987comparison}.}

{By contrast, the PRS approach replaces the point null with a composite null hypothesis. Rather than asking whether \emph{any} change has occurred, it asks whether the change exceeds a pre-specified, tolerable threshold. This distinction matters. It separates practical and statistical significance, grounding the test in business relevance rather than arbitrary $p$-values. It also enables clearer communication with non-technical stakeholders, as decision thresholds are now expressed in intuitive terms, (e.g., ``a shift of more than 5\% in any risk category.'')}

{Finally, the PRS framework supports a more nuanced, three-tiered decision process. By using nested composite hypotheses, it formalizes the familiar ``red–amber–green'' (RAG) structure in a way that is statistically coherent, flexible, and easily aligned with institutional risk appetite. We also emphasize that the specification of a tolerable shift is not unconstrained: its scale must respect structural limits imposed by the sample size and number of categories. These constraints, and a practical tool for calibration, are presented in Section~3.4.}

\subsection{Research aims and objectives}

In light of the methodological contributions outlined above, this paper sets out to establish the PRS as a principled and practical alternative to the PSI for monitoring distributional shifts. The PRS framework incorporates explicit risk tolerances, adapts to sample size, and supports decision-making through nested composite hypotheses. The key objectives of this research are as follows:

\vspace{-0.5em}
\begin{itemize}
    \item To demonstrate the limitations of the PSI in its current form, {including its sensitivity to minor deviations and challenges in various portfolio sizes}.
    \item To introduce $\delta$-resemblance as a practical, interpretable way to bound population shift and inform business decisions.
    \item To develop the PRS as an alternative monitoring tool with well-calibrated critical values.
    \item To formulate the PRS under a composite null hypothesis, relaxing the rigidity of a point null, and justify decision thresholds via the least favorable non-central chi-square distribution.
        \item To provide a practical framework for implementing the PRS in credit risk modeling and other areas where population monitoring is a concern.
    \item To evaluate PRS performance through simulation studies and real-world applications, {highlighting its competency in both small and large sample sizes}.

\end{itemize}


This remainder of the paper is structured as follows: Section \ref{sec: measuring resemblance} formalizes the problem, introducing the PRS as an alternative discrepancy measure with a brief discussion of related methods. Section \ref{sec: PRS properties} outlines the statistical properties of the PRS, while Section \ref{sec:framework} derives the sample-size dependent decision boundaries (critical values). Section \ref{sec:implementation} contains a practical guide to implementing the PRS, including a flowchart for application. Section \ref{sec:PRS simulation} presents a comprehensive simulation study, and Section \ref{Application} applies the PRS to real-world data, comparing it with the widely used PSI and discrete Kolmogorov-Smirnov (KS) statistic. Finally, Section \ref{sec:Conclusion} summarizes key contributions and suggests directions for future research.

\section{Measuring population resemblance}
\label{sec: measuring resemblance}

\subsection{Statistical representation}

Consider an independent and identically distributed (i.i.d.) sample of ordinal scores, $X_1, X_2, \ldots, X_n$, from a population with cumulative distribution function (cdf) $F$ defined on the set of integers $\{1,2,\ldots,B\}$. These scores often arise by discretizing numerical values with each score representing membership in one of $B\geq 2$ disjoint categories (e.g. level of risk). {In practice, these categories do not only reflect levels or risk, but are often also determined by operational considerations such as Basel-compliant rating grades, product segmentation or business lines, rather than through formal statistical procedures. As such, the number of categories $B$ are considered fixed a priori.} 

Let $n_i$ denote the count of scores in category $i$, formally expressed as $n_i = \sum_{j=1}^{n} \mathbb{I}(X_j = i)$, where $\mathbb{I}(A)$ is the indicator function such that $\mathbb{I}(A)=1$ when $A$ is true and $\mathbb{I}(A)=0$ otherwise. The total number of observed scores, $n$, is equal to the sum of all category counts, $n = \sum_{i=1}^{B} n_i$. 

The true probability of a score falling into category $i$ is $p_{i} = P(X_j = i) = F(i)-F(i-1)$ for $i=1,\ldots,B$.  Let $\bm{p}=(p_1,p_2,\ldots,p_B)^\top$. The observed category proportions, serving as unbiased estimators of the true probabilities, are denoted by $\widehat{\bm{p}} = (\hat{p}_1, \hat{p}_2, \ldots, \hat{p}_B)^\top$ with $\hat{p}_i = n_i/n$ for $i = 1, 2, \ldots, B$. 

The aim is to determine whether the current population $\bm{p}$ is sufficiently close to the reference population $\bm{p}_0 = (p_{01}, p_{02}, \ldots, p_{0B})^\top$ on which the model was built. The specific nature of this model is not germane. Rather, the question is whether $\bm{p}$ has “substantively” shifted from $\bm{p}_0$ or still “resembles” it. While the current population $\bm{p}$ is unknown in practice, $\widehat{\bm{p}}$ serves as an unbiased estimator. {We assume that the reference population probabilities satisfy $p_{0i}> 0$ for all $i=1,\ldots,B$, ensuring that each category in the reference population has a non-zero probability of occurrence. Additionally, the estimated probabilities, derived from the observed data, are non-negative, i.e. $\hat{p}_{i}\geq 0$ for all $i=1,...,B$.} Furthermore, the random variable $n\widehat{\bm{p}}$ follows a multinomial distribution with expectation $\mathrm{E}[n\widehat{\bm{p}}\,] = n\bm{p}$ and covariance matrix $\mathrm{Var}[n\widehat{\bm{p}}\,] = n\big[\mathrm{diag}(\bm{p})-\bm{p}\bm{p}^\top\big].$ Here, $\mathrm{diag}(\bm{p})$ is a diagonal matrix with the elements of $\bm{p}$ on the diagonal.

\subsection{Existing measures of population resemblance}

Several measures have been proposed for quantifying population shift in credit risk modeling. The most widely used is the PSI, introduced by \textcite{lewis1994introduction}. Defined as
\begin{equation}
    \mathrm{PSI} = \sum_{j=1}^{B} (\hat{p}_{j} - p_{0j})(\log{\hat{p}_j} - \log{p_{0j}}) \mathbb{I}(\hat{p}_{j} > 0), \label{eq:PSI}
\end{equation}
the PSI serves as a consistent estimator of the symmetric Kullback-Leibler divergence
\begin{equation}
\label{eqn:J}
J := J(\bm{p}, \bm{p}_0) = \sum_{j=1}^{B} ({p}_j - {p}_{0j})(\log{{p}_j} - \log{{p}_{0j}}).
\end{equation}
This $ J $, first introduced by \textcite{jeffreys1948theory}, is a symmetrized version that addresses the inherent asymmetry of the original ``expected per observation information,'' here equivalent to 
$
I(\bm{p}, \bm{p}_0) = \sum_{j=1}^{B} p_j (\log{{p}_j} - \log{{p}_{0j}})$, see \textcite{pichler2020entropy}.

Other measures in risk modeling, {inheriting their properties from the Kullback-Leibler diverge,} include the Information Value \parencite[p.~184]{siddiqi2017intelligent} and the Characteristic Stability Index \parencite[p.~369]{siddiqi2017intelligent}. Like the PSI, these measures rely on arbitrary thresholds that disregard sample size and statistical properties.

The chi-square divergence, 
\begin{equation} \chi^2 := \chi^2(\bm{p}, \bm{p}_0) = \sum_{j=1}^{B} \frac{(p_j - p_{0j})^2}{p_{0j}}, \label{eq:chisq_div} \end{equation}
offers a statistically principled alternative for measuring population differences. Both $J$ and $\chi^2$ are members of the $f$-divergence family \parencite{renyi1961measures,csiszar1967information}, with $\chi^2$ providing a local approximation to $J$ (and other $f$-divergences) when populations are similar \parencite{csiszar2004information}. This property, combined with its well-understood statistical behavior, makes $\chi^2$ an attractive foundation for population monitoring.

Building on these advantages, we propose the PRS, defined as 
\begin{equation}
    \mathrm{PRS} = \sum_{j=1}^{B} \frac{(\hat{p}_{j} - p_{0j})^2}{p_{0j}}, \label{eq:PRS}
\end{equation}
which offers a measure grounded in well-established statistical theory and extensively studied properties under a broad range of conditions, thereby also addressing key limitations of the PSI. Specifically, we will develop methods to incorporate sample-size-dependent {critical values}, using the limiting non-central $\chi^2$ properties of the normed PRS to provide a principled framework for evaluating population shifts.

Other measures include the Kolmogorov-Smirnov statistic \parencite{dagostino1986goodness}, $\mathrm{KS}=\max_{j=1,...,B} \big|\hat{F}(j)-F_0(j)\big|$, where $\hat{F}(j) = \sum_{i=1}^{j} \hat{p}_i$ is the empirical cdf and $F_0(j)=\sum_{i=1}^{j}p_{0i}$ is the cdf of the model construction population. However, the KS statistic has limited utility for discrete distributions, as its (asymptotic) distribution --- unlike in the continuous setting --- depends on the underlying $\bm{p}_0$ \parencite{conover1972kolmogorov}, making critical values less straightforward to determine. {The \texttt{R} package \texttt{dgof} offers an implementation of the KS test tailored for discrete settings \parencite{RDGOF}.} The more recent statistic of \textcite{dupisanievisagie2020}, $\mathrm{DPV}=\max_{j=1,\ldots,B^\ast}|\hat{p}_j - p_{0j}| / p_{0j}$, relies on an arbitrary selection of $B^\ast < B$ and Monte Carlo calibration. Its use has been explored only in settings with $n \geq 10{,}000$. For further reading on the discrete goodness-of-fit problem, see \textcite{agresti2012categorical}.

\subsection{Population resemblance framework}
\label{sec:PR Framework}

To formalize the assessment of population shift, we introduce the concept of $\delta$-resemblance, which quantifies acceptable deviations between probability distributions and provide a structured framework for analyzing the behavior of the PRS.

\begin{definition}\label{deltaresemblance}
Let $\delta>0$. For the probability vector $\bm{p}_0$, define the region  
$$
\mathcal{P}(\delta|\bm{p}_0) = \Bigg\{\tilde{\bm{p}} = (\tilde{p}_1,\ldots,\tilde{p}_B) : \max_{j=1,\ldots,B} |p_{0j} - \tilde{p}_j| \leq \delta,\ \sum_{j=1}^{B}\tilde{p}_j = 1 \Bigg\}. 
$$
The probability vector $\bm{p}$ is said to be $\delta$-resemblant of $\bm{p}_0$ whenever $\bm{p}\in \mathcal{P}(\delta|\bm{p}_0)$.
\end{definition}

Intuitively, $\mathcal{P}(\delta|\bm{p}_0)$ defines the set of all valid probability vectors where no category probability $p_j$ deviates from its reference value, $p_{0j}$, by more than $\delta$. This tolerance-based approach ensures that deviations remain manageable and within acceptable limits, making it an ideal framework for practical applications.

This paper’s methodology builds on $\delta$-resemblance and supporting technical conditions to ensure accurate application. Specifically, we assume that the reference probabilities are strictly positive ($p_{0j} > 0$ for all $j = 1, \ldots, B$) to ensure that every category in the reference distribution is represented. Furthermore, we impose the constraint $0 < \delta \leq \min\limits_{j=1, \ldots, B} p_{0j}$, which guarantees that the parameter $\delta$ does not exceed the smallest reference probabilities. This condition effectively limits the maximum allowable shift for any single category to its own probability mass. Finally, $\delta$ is designed to scale with both the sample size $n$ and the number of categories $B$, allowing it to adapt to variability across different portfolio structures and data conditions. For a given implementation, it is a pre-specified parameter, not estimated from the data, reflecting the organization’s risk tolerance (i.e., the maximum acceptable deviation from the reference population and ensures consistency in decision-making{)}.

The formulation of Definition~\ref{deltaresemblance} leverages the Chebyshev distance {(also known as the $\ell_\infty$ norm)} to define a region of tolerable shift, by bounding the largest absolute difference between corresponding category probabilities. This distance is particularly advantageous in risk-sensitive applications due to its direct interpretability and provides a precise, actionable metric for monitoring changes in probability distributions. {The Chebyshev distance aligns with the operational need to control critical deviations, as it isolates the largest shift in the data. This characteristic is especially important in credit risk monitoring, where changes in high-risk categories may have outsized implications. Unlike alternatives such as the Euclidean distance, which averages deviations across categories, the Chebyshev distance focuses on the worst-case deviation, offering sharper insights into localized shifts.}

{In a credit risk management application, consider the portfolio-level probability of default (PD), computed as $\mathrm{PD}(\bm{{p}})=\left(\sum_{i=1}^B n_i\, \mathrm{pd}_i\right) / \sum_{i=1}^B n_i =\sum_{i=1}^B (n_i/n)\, \mathrm{pd}_i = \sum_{i=1}^Bp_i\, \mathrm{pd}_i$ where $n_i$ denotes the number of borrowers and $\mathrm{pd}_i$ the default probability in each risk category $i$. For two $\delta$-resemblant populations $\bm{p}$ and $\bm{p'}$, the difference in their portfolio-level PD is bounded by $|\mathrm{PD}(\bm{p})-\mathrm{PD}(\bm{p'})|\leq \delta\sum_{i=1}^B \mathrm{pd}_i$.} This result shows that the $\delta$-resemblance framework balances statistical rigor with practical relevance, enabling risk managers to quantify the population shift’s impact on portfolio performance, assess acceptable levels of deviation aligned with business risk tolerance, and decide when model recalibration is warranted.

{We note that a two-sample formulation of the problem, i.e. treating $\bm{p}_0$ as arising from a random sample from the model construction population, may have some appeal. In practice, however, this approach is complicated by dependencies between the model construction and current samples, due to temporal evolution or data overlap. To address the issue of non-independence between the populations $\bm{p}_0$ and $\bm{p}$, we have thus resorted to a one-sample problem formulation, treating the model construction probabilities $\bm{p}_0$ as fixed and known. Even in situations as described where the model construction probabilities were established using sampling tools, the one-sample approach can still be employed, providing a conditional inference solution.} 


\section{Statistical properties of Population Resemblance}
\label{sec: PRS properties}

\subsection{Small sample behavior}
\label{sec:small_sample}

{In this subsection, we investigate the finite sample properties of the PSI and PRS, examining both scenarios where the current population $\bm{p}$ matches and departs from the reference population $\bm{p}_0$ used in model construction. Critically, $\bm{p}_0$ is assumed to be fixed and known throughout the paper.} Define the scaled statistics $T_n = n\times \mathrm{PSI}$ and $Q_n = n\times \mathrm{PRS}$ where $\mathrm{PSI}$ and $\mathrm{PRS}$ are given in \eqref{eq:PSI} and \eqref{eq:PRS}, respectively. Both of these are known to follow a (central) $\chi^2$ distribution with $B-1$ degrees of freedom under the assumption of no population shift, $\bm{p}=\bm{p}_0$, as the sample size increases. However, in smaller samples, their behavior can deviate significantly from asymptotic expectations, making it important to understand these differences for practical applications.

Despite the well-established asymptotic distribution of $T_n$ \parencite[Chapter~6]{kullback1978information}, practitioners often rely on the thresholds of \textcite{lewis1994introduction}, which prescribe $\mathrm{PSI} < 0.1$ to indicate acceptable population similarity and $\mathrm{PSI} \geq 0.25$ as a trigger for model reconstruction. These thresholds do not account for critical factors such as sample size or the number of categories, which can substantially affect the behavior and interpretation of the PSI in finite-sample settings.

{To illustrate the limitations of fixed thresholds, we conducted a simulation study examining the probability of mandating model reconstruction under two scenarios:
\begin{enumerate}
\item No population shift: $\bm{p} = \bm{p}_0$.
\item Moderate shift: $\bm{p}$ differs from $\bm{p}_0$ such that $J(\bm{p}, \bm{p}_0) = 0.1$, where $\bm{p}$ was determined by minimizing the Euclidean distance between $\bm{p}$ and $\bm{p}_0$ subject to a Lagrange multiplier constraint and the simplex constraint.
\end{enumerate}
Both scenarios assume equal model construction probabilities $\bm{p}_0 = (1/B, \ldots, 1/B)$. The probabilities $\hat{\bm{p}}$ used for calculating the $\mathrm{PSI}$ were obtained by scaling multinomial counts simulated with $n$ ranging from 50 to 500. Table \ref{TABLE1} presents the estimated probabilities of mandating reconstruction, $\widehat{P}(\mathrm{PSI} \geq 0.25)$, where $\widehat{P}$ represents the estimated probability based on $K = 10^6$ simulated datasets.}

The simulation results reveal two critical issues with fixed PSI thresholds: For small samples ($n = 50$), reconstruction is mandated too frequently, even under no shift. For larger samples ($n = 500$), reconstruction is rarely triggered, even with moderate shifts.

\begin{table}[htbp]
    \centering
    \caption{Estimated probabilities of mandating reconstruction under fixed PSI thresholds.}
    \label{TABLE1}
    \begin{tabular}{p{1cm}p{2cm}p{2cm}p{2cm}p{2cm}}
        \toprule
         & \multicolumn{2}{c}{$J = 0$} & \multicolumn{2}{c}{$J = 0.1$} \\
        \cmidrule(lr){2-3} \cmidrule(lr){4-5}
        $n$& $B = 5$ & $B = 10$ & $B = 5$ & $B = 10$ \\
        \cmidrule(lr){1-5} 
        50  & 0.0226 & 0.2356 & 0.2542 & 0.5459 \\
        100 & 0.0001 & 0.0086 & 0.0872 & 0.2508 \\
        200 & 0.0000 & 0.0000 & 0.0131 & 0.0434 \\
        500 & 0.0000 & 0.0000 & 0.0001 & 0.0003 \\
        \cmidrule(lr){1-5} 
    \end{tabular}
\end{table}

\begin{figure}[h!]
	\centering
	\begin{minipage}{0.48\textwidth}
		\centering
\includegraphics[trim={0.15cm 5cm 18cm 0cm},clip,width=0.6\textwidth,angle=90]{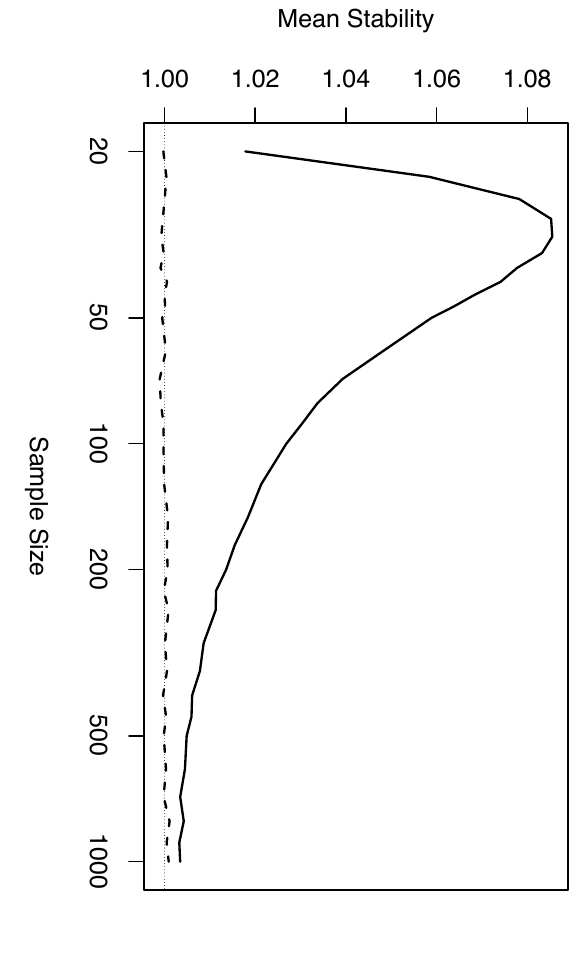} 
		\caption{Mean stability ratios for $T_n$ (solid) and $Q_n$ (dashed).}
        \label{fig:mean_stab_ratios}
	\end{minipage}\hfill
	\begin{minipage}{0.48\textwidth}
		\centering
\includegraphics[trim={0.15cm 5cm 18cm 0cm},clip,width=0.6\textwidth,angle=90]{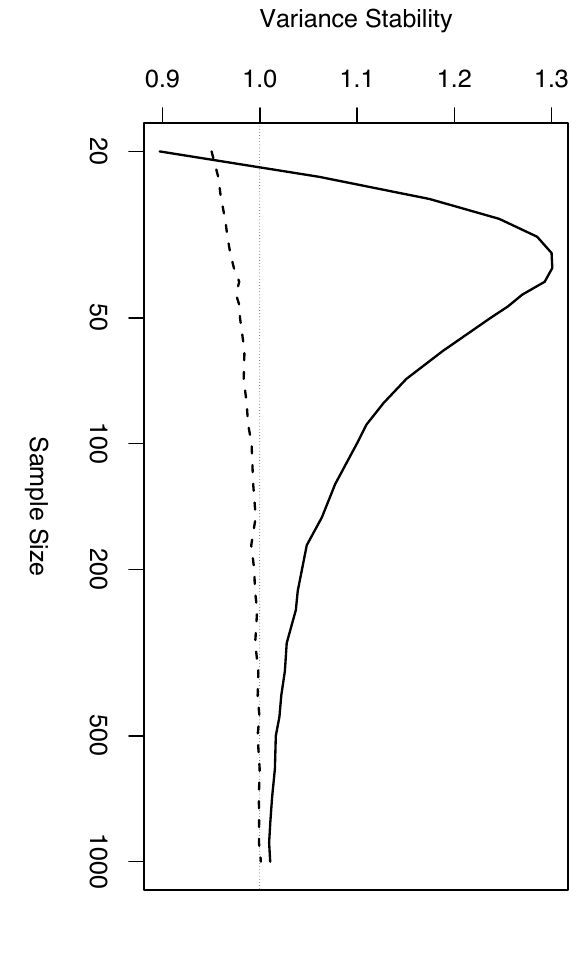} 
		\caption{Variance stability ratios for $T_n$ (solid) and $Q_n$ (dashed).}
        \label{fig:var_stab_ratios}
	\end{minipage}
\end{figure}

To better understand these behaviors, we examine the mean and variance stability of the statistics. For $T_n$, define the stability ratios,
$$
\Lambda^{(1)}_{T_n} = \frac{\mathrm{E}[T_n]}{B-1} \qquad \text{and} \qquad \Lambda^{(2)}_{T_n} = \frac{\mathrm{Var}[T_n]}{2(B-1)}.
$$
The corresponding quantities for $Q_n$ are defined similarly. These ratios measure convergence to asymptotic moments, with values near 1 indicating stability. Using $K = 10^6$ Monte Carlo realizations for sample sizes from 20 to 1{,}000, we estimate these ratios for $B=5$ categories. Figures \ref{fig:mean_stab_ratios} and \ref{fig:var_stab_ratios} display the results.

After appropriate adjustment for sample size, the PRS (as per $Q_n$) exhibits mean stability even at small sample sizes and variance stability for $n \geq 50$. In contrast, the PSI (as per $T_n$) shows substantial instability, with empirical moments deviating from asymptotic values by up to 8\% (mean) and 30\% (variance) in small samples. Although close agreement between empirical and asymptotic moments does not ensure convergence, substantial discrepancies indicate a lack of asymptotic convergence at the corresponding sample size.

\subsection{Non-central chi-square limiting distribution}
\label{sec:limiting}

Having established the superior small-sample properties of the PRS under no population shift in comparison to the PSI, we now examine PRS behavior under population shift, where the current population $\bm{p}$ may differ from the reference population $\bm{p}_0$ by up to a specified tolerance $\delta$. Recall the sample-size normalized PRS statistic, {\[
Q_n = n\times \mathrm{PRS} = n \sum_{j=1}^{B} \frac{(\hat{p}_j - p_{0j})^2}{p_{0j}}.\]}Let $\delta_j = p_j - p_{0j}$ denote the category-specific deviations, assumed to be small and of the order $n^{-1/2}$; that is, $p_j-p_{0j}=\text{O}(n^{-1/2})$. Formally, we assume there exist constants $\xi_j$ such that $\lim_{n\rightarrow\infty} n^{1/2}\delta_j = \xi_j.$ Thus, we write $p_j = p_{0j} + \delta_j$, where $\bm{p}$ is a so-called \textit{local alternative} near $\bm{p}_0$.

{Under multinomial sampling, the count vector $ (n_1, \ldots, n_B)^\top = n\times (\hat{p}_1,\ldots,\hat{p}_B)^\top$ follows
a $\text{Multinomial}(n, \bm{p})$ distribution. By the multivariate central limit theorem for proportions, we have
\[
\sqrt{n}(\widehat{\bm{p}} - \bm{p}) \xrightarrow{d} \text{N}\big(0, \bm{\Sigma}(\bm{p})\big)
,
\]
where $\bm{\Sigma}(\bm{p}) = \mathrm{diag}(\bm{p}) - \bm{p} \bm{p}^\top$ is a singular $B \times B$ covariance matrix of rank $B-1$, reflecting the constraint that the probabilities sum to one.
}


{Expanding $Q_n$ using the decomposition $\hat{p}_j - p_{0j} = (\hat{p}_j - p_j) + (p_j - p_{0j})$, and letting $\bm{\delta} = \bm{p} - \bm{p}_0$ and $\mathbf{W} = \mathrm{diag}(\bm{p}_0)$, gives
\begin{eqnarray*}
Q_n &=& \sum_{j=1}^{B} \left[ \frac{n(\hat{p}_j - p_j)^2}{p_{0j}} + \frac{2n(\hat{p}_j - p_j)(p_j - p_{0j})}{p_{0j}} + \frac{n(p_j - p_{0j})^2}{p_{0j}} \right] \\
&=& n(\widehat{\bm{p}} - \bm{p})^\top \mathbf{W}^{-1} (\widehat{\bm{p}} - \bm{p}) 
+ 2n\bm{\delta}^\top \mathbf{W}^{-1}(\widehat{\bm{p}} - \bm{p}) 
+ n\bm{\delta}^\top \mathbf{W}^{-1} \bm{\delta}.
\end{eqnarray*}
Above, the first term converges to a chi-square distribution with $B-1$ degrees of freedom, $n(\widehat{\bm{p}} - \bm{p})^\top \mathbf{W}^{-1} (\widehat{\bm{p}} - \bm{p}) \xrightarrow{d} \chi^2_{B-1}.$
The second term is linear in the centered random vector and satisfies
 $2n\bm{\delta}^\top \mathbf{W}^{-1}(\widehat{\bm{p}} - \bm{p}) = 2\bm{\xi}^\top \mathbf{W}^{-1} \sqrt{n}(\widehat{\bm{p}} - \bm{p}) + \text{o}_p(1),$
which converges in distribution to a normal random variable. Here, $\text{o}_p(1)$ denotes a remainder term that converges to zero in probability. The third term is deterministic and forms the non-centrality parameter, $\lambda = n \bm{\delta}^\top \mathbf{W}^{-1} \bm{\delta} = \sum_{j=1}^{B} n\delta_j^2/p_{0j}.$ Combining these results and applying standard theory on quadratic forms under local alternatives (\cite{cressie1984multinomial}), we obtain the following limiting distribution.}

\begin{proposition}
Assume $\delta_j=p_j - p_{0j} = \text{O}(n^{-1/2})$ for $j=1,\ldots,B$. Then, as $n \to \infty$,
\vspace{-5mm}
\[
Q_n \overset{d}{\longrightarrow} \chi^2_{B-1}(\lambda),
\]

\vspace{-5mm}
\noindent a non-central chi-square distribution with $B-1$ degrees of freedom and non-centrality parameter $\lambda$ given by
\[
\lambda = \sum_{j=1}^B \frac{n\delta_j^2}{p_{0j}} = \sum_{j=1}^B \frac{n(p_j - p_{0j})^2}{p_{0j}}.
\]
\end{proposition}\bigskip

Since the true values of $p_j$ are typically unknown, direct evaluation of $\lambda$ is infeasible. Moreover, while the definition of $\lambda$ suggests a dependence on sample size $n$, it remains well-defined in the asymptotic framework due to the assumed $n^{-1/2}$ decay in the $\delta_j$.

To address the challenge of $\lambda$ being unknown, we adopt a conservative approach by framing the problem in terms of the maximal non-centrality parameter under $\delta$-resemblance. This aligns with the concept of least favorable distributions, where test statistics are evaluated under the worst-case scenario within the null hypothesis, see \textcite{reinhardt1961use}. This approach allows us to construct robust decision-making critical values using the supremum of $\lambda$, defined as  

\vspace{-2mm}
\[
\lambda_{\mathrm{sup}} = \sup_{\bm{p} \in \mathcal{P}(\delta | \bm{p}_0)} \sum_{j=1}^B \frac{n\,(p_j - p_{0j})^2}{p_{0j}}.
\]

\vspace{2mm} A key property of the non-central chi-square distribution underpins this framework: stochastic ordering. For $\lambda \leq \lambda_{\mathrm{sup}}$ with fixed degrees of freedom $B-1$,
\vspace{-5mm}
\[
\chi^2_{B-1}(\lambda_{\mathrm{sup}}) \preceq_{\mathrm{st}} \chi^2_{B-1}(\lambda),
\]

\vspace{-5mm}
\noindent where $\preceq_{\mathrm{st}}$ denotes stochastic dominance. That is, if $X \sim \chi^2_{B-1}(\lambda_{\mathrm{sup}})$ and $Y \sim \chi^2_{B-1}(\lambda)$, the cumulative distribution functions satisfy $F_Y(x) \geq F_X(x) \text{ for all } x \in \mathbb{R}.$ This property ensures that the maximal non-centrality parameter $\lambda_{\mathrm{sup}}$ provides a conservative basis for decision-making, enabling robust inference despite uncertainty about the true $\bm{p}$.

In the next proposition, we establish how the maximal non-centrality parameter $ \lambda_{\mathrm{sup}} $ depends on the reference probabilities $ \bm{p}_0 $ and the partition size $ B $, providing a precise characterization of $ \lambda_{\mathrm{sup}} $ in terms of these parameters.

\begin{proposition}
Under a constraint of $\bm{p}\in \mathcal{P}(\delta|\bm{p}_0)$, the maximal non-centrality parameter is
\[
\lambda_{\mathrm{sup}} =
\begin{cases}
    n\delta^2 \displaystyle\sum_{j=1}^B p_{0j}^{-1}, & \text{if } B \text{ is even,} \\
    n\delta^2 \left( \displaystyle\sum_{j=1}^B p_{0j}^{-1} - p_{\ast}^{-1} \right), & \text{if } B \text{ is odd,}
\end{cases}
\]
where $p_{\ast} = \max\limits_{j=1,\ldots,B} p_{0j}$.
\end{proposition}

\begin{proof}
To derive the maximal non-centrality parameter $ \lambda_{\mathrm{sup}} $, we consider the supremum of $ \lambda $ under the constraint of $ \delta $-resemblance, $ \bm{p} \in \mathcal{P}(\delta \,|\, \bm{p}_0) $. The convexity of $ \lambda $ as a function of $ \bm{p} $ ensures that its maximum is attained at an extreme point of the feasible set $ \mathcal{P}(\delta \,|\, \bm{p}_0) $. Translating $ \mathcal{P}(\delta \,|\, \bm{p}_0) $ by $ \bm{p}_0 $, {this set becomes the intersection of a $ \delta $-scaled Chebyshev ball (i.e., a hypercube under the $\ell_\infty$ norm) and a codimension-1 vector subspace of $\mathbb{R}^B$ that passes through the center of the hypercube and is orthogonal to the main diagonal.} Each extreme point of this set has coordinates $ p_j \in \{ p_{0j} - \delta,\, p_{0j},\, p_{0j} + \delta \} $, with at most one $ p_j $ remaining at $ p_{0j} $, and with the number of $ +\delta $ and $ -\delta $ deviations being equal. Two distinct cases need to be considered.

\textit{Case 1}: For $ B $ even, no coordinate remains unperturbed, meaning all components $ p_j $ take values $ p_{0j} \pm \delta $. Consequently, all extreme points yield the same value of $ \lambda $, and we find
\[
\lambda_{\mathrm{sup}} = n\delta^2 \sum_{j=1}^B p_{0j}^{-1}.
\]

\textit{Case 2}: For $ B $ odd, symmetry requires that one coordinate $ p_j $ remains unperturbed ($ p_j = p_{0j} $). To maximize $ \lambda $, this zero increment is assigned to the index $ j $ corresponding to the largest $ p_{0j} $, as this minimizes the term $ 1 / p_{0j} $. Substituting this condition, the supremum becomes
\[
\lambda_{\mathrm{sup}} = n\delta^2 \left( \sum_{j=1}^B p_{0j}^{-1} - p_{\ast}^{-1} \right),
\]
where $ p_{\ast} = \max\limits_{j=1,\ldots,B} p_{0j} $.

Thus, $ \lambda_{\mathrm{sup}} $ is now fully characterized for all $ B $, completing the proof.
\end{proof}

\subsection{{Decision-making framework for population monitoring}}
\label{sec:framework}

{As previewed in Section~\ref{sec:contrib}, the PRS framework replaces the traditional point null with a pair of nested composite hypotheses, enabling a three-tiered response structure that distinguishes tolerable from material shifts and supports practical risk management. This section formalizes the hypotheses, defines decision regions, and explains the operational role of tuning parameters.}

With thanks to a reviewer for identifying this connection, we formalize our initial proposal using a nested hypothesis-testing framework based on the concept of $\delta$-resemblance. Let $H_{01} : \bm{p} \in \mathcal{P}(\delta \mid \bm{p}_0)$ denote the hypothesis that the current population resembles the reference population $\bm{p}_0$ up to a maximum deviation of $\delta$, as defined in Section~\ref{sec: measuring resemblance}. Let $H_{02} : \bm{p} \in \mathcal{P}(M\delta \mid \bm{p}_0)$ denote a more relaxed hypothesis allowing deviation up to $M\delta$, with $M > 1$ a user-specified multiplier. These hypotheses are nested, with $H_{01} \subset H_{02}$.

This structure supports a three-level decision process, determined by which hypotheses are rejected. If neither $H_{01}$ nor $H_{02}$ is rejected, the model is considered acceptable for continued use. If $H_{01}$ is rejected but $H_{02}$ is not, the population shift exceeds the threshold of $\delta$ but remains within the bounds of $M\delta$, prompting enhanced monitoring. If both hypotheses are rejected, the shift is deemed substantial enough to warrant full model reconstruction.

\begin{definition}
\label{DefRegions}
The PRS decision regions are defined as follows,
\vspace{-5mm}
\begin{align*}
\mathcal{R}_1 &= \left\{\mathrm{PRS} \leq \tau_1 \right\} 
&&\text{(Continue using model, ``acceptable'')}\\
\mathcal{R}_2 &= \left\{\tau_1 < \mathrm{PRS} \leq\tau_2\right\} 
&&\text{(Enhanced monitoring, ``partially discrepant'')}\\
\mathcal{R}_3 &= \left\{\mathrm{PRS} > \tau_2 \right\}  && \text{(Reconstruct model, ``fully discrepant'').}
\end{align*}

\vspace{-5mm}
\noindent These {decision} boundaries, or ``critical values'', are defined as 
\vspace{-1mm}
$$\tau_1:=\frac{F_{B-1}^{-1}(\alpha_2,\,M^2\lambda_{\mathrm{sup}})}{n} \ \ \ \text{and} \ \ \ \tau_2:=\frac{F_{B-1}^{-1}(1-\alpha_1,\,\lambda_{\mathrm{sup}})}{n},$$ 

\vspace{-5mm}
\noindent where $0 \leq \alpha_1, \alpha_2 \leq 1$ reflect the organization’s risk tolerance for erroneous decisions, and $F_{B-1}^{-1}(\alpha, \lambda)$ denotes the $\alpha$-quantile of the non-central $\chi^2$ distribution with $B - 1$ degrees of freedom and non-centrality parameter $\lambda$.
\end{definition}

{The labels ``acceptable'', ``partially'' and ``fully discrepant'' are illustrative and not industry agnostic. In banking, for example, a red-amber-green ($\mathbbm{RAG}$) classification is commonly used to describe degrees of discrepancy. Then,  $\mathbbm{R}$ indicates membership to $\mathcal{R}_3$, $\mathbbm{A}$ to $\mathcal{R}_2$ and $\mathbbm{G}$ to $\mathcal{R}_1$. These descriptive labels should be tailored to the application context. Alternatives might include ``low, medium, high risk of deviance.''} 

\begin{figure}[h]
    \centering
    \includegraphics[trim={0 0.58cm 0.1cm 0.3cm},clip,height=5cm]{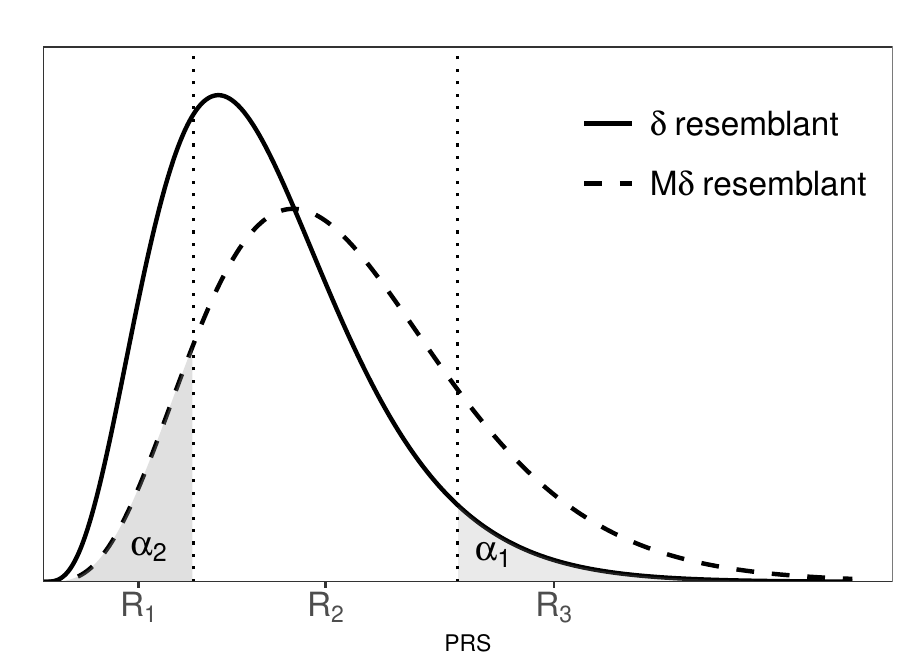}
    \caption{Schematic representation of PRS decision-making critical values.}
    \label{fig:three_regions}
\end{figure}

Within this framework, $\alpha_1$ serves as a Type I error rate under $H_{01}$, determining the likelihood of rejecting both $H_{01}$ and $H_{02}$ when the population is (only) $\delta$-resemblant. This corresponds to incorrectly concluding that the population is fully discrepant, leading to model reconstruction. As such, $\alpha_1$ defines the decision boundary for crossing from $\mathcal{R}_2$ to $\mathcal{R}_3$ and governs the corresponding upper critical value $\tau_2$ in Definition~\ref{DefRegions}. Conversely, $1-\alpha_2$ is the likelihood of exiting $\mathcal{R}_1$ when the population is at the boundary of $M\delta$-resemblance. While this resembles a detection of power, it is still defined within the null hypothesis framework of $H_{02}$ rather than, as conventional power, against an alternative hypothesis. Instead, $\alpha_2$ controls the likelihood of maintaining the current model despite a potential shift of size $M\delta$. Thus, it defines the decision boundary for crossing from $\mathcal{R}_1$ to $\mathcal{R}_2$ --- the lower critical value $\tau_1$ in Definition~\ref{DefRegions}. The decision regions are schematically illustrated in Figure \ref{fig:three_regions} that depicts the PRS density curves derived from the non-central $\chi^2$ distribution. Tail regions corresponding to $\alpha_1$ and $\alpha_2$ are highlighted.

Using two null hypotheses for the same parameter is rare but not unprecedented; \textcite{schuirmann1987comparison}, for example, applies a related strategy in bioequivalence testing. {The use of two nested composite null hypotheses is what permits $\alpha_1$ and $\alpha_2$ to be specified independently, each tied to a distinct decision boundary with separate practical implications. Unlike traditional approaches that apply two significance levels to a single null hypothesis, the nested structure separates the risks of overreaction (rejecting a tolerable deviation) from underreaction (failing to act on a material one). This separation supports more meaningful discussions with stakeholders, as the two types of monitoring errors can be aligned with different business consequences and risk tolerances.} Moreover, this structure, informed by the stochastic dominance of the non-central chi-square distribution, upholds a conservative decision-making process while allowing practitioners to fine-tune decision sensitivity and maintain clear boundaries for monitoring and intervention.

{While the stochastic ordering result in Section~\ref{sec:limiting} ensures robust Type I error control, it does not extend to settings where $\bm{p} \notin \mathcal{P}(M\delta \mid \bm{p}_0)$. In such cases, evaluating $ \mathbb{P}(\mathrm{PRS} \in \mathcal{R}_j) $ for $j = 1, 2, 3$ requires either specifying a concrete alternative or working within a least favorable alternative framework. These approaches introduce stronger modeling assumptions and greater technical complexity, which fall outside the scope of this paper’s focus on error calibration. Developing such power-based refinements is an important direction for future research.}


\subsection{Implementation guide}
\label{sec:implementation}

{Implementing the PRS framework in practice requires a principled approach, especially in parameter selection. This section offers practical guidance for risk managers and analysts to ensure valid and effective implementation of the PRS across risk portfolios of varying sizes. Central is the tolerable shift parameter $\delta$, which defines the maximum acceptable category-wise deviation. This value is not freely specifiable: it must respect theoretical constraints imposed by the sample size $n$ and the number of categories $B$, ensuring the validity of the underlying non-central chi-square approximation.} 

Motivated by the asymptotic theory, we recommend defining $\delta$ as
\begin{equation}
\delta = c\ \times \min_{j=1,\ldots,B} \left\{ \frac{p_{0j}(1-p_{0j})}{n} \right\}^{1/2},
\label{eqn:rec_delta}
\end{equation}
where $c>0$ is a scaling factor that adjusts the acceptable magnitude of shift in relation to sampling variability. Since the quantity $\{p_{0j}(1-p_{0j})/n\}^{1/2}$ corresponds to the standard error of a sample proportion, it provides a natural benchmark for defining what constitutes a meaningful deviation. In this setup, $c$ functions like a tolerance multiplier --- comparable to the number of standard errors deemed acceptable --- supporting consistent application across portfolios with different sizes and risk profiles.

For example, in a credit scoring context, $\delta$ limits category-specific shifts (e.g., a rise in high-risk borrowers) that would undermine confidence in the model's continued validity. The PRS is calibrated to detect when one or more categories exhibit such levels of shift. Tying $\delta$ to sampling variability guards against substantive shifts across categories while maintaining the connection to sample size and category-specific uncertainty. 

While examples in later sections assume an equi-probable $\bm{p}_0$ for simplicity, the PRS applies to arbitrary reference distributions. Importantly, although the PRS statistic is asymptotically distribution-free, the decision boundaries defined using $\delta$ in \eqref{eqn:rec_delta} do depend on $\bm{p}_0$. The recommendation in \eqref{eqn:rec_delta} is but one possible solution to appropriately adjust for sample size. {However, this flexibility is moderated by the mathematical constraint that $M\delta \leq \min_j p_{0j}$. This ensures that perturbed category probabilities remain valid. }

{As a result, for highly imbalanced reference distributions (i.e., when some $p_{0j}$ are small), the allowable range of $\delta$ may become narrow, limiting the magnitude of detectable shifts.} The parameter $M>1$, which controls the threshold between partial and full discrepancy, should be chosen accordingly. Smaller values of $M$ are appropriate in settings that demand rapid response, while larger values allow for more tolerance before action is triggered. Care should be taken, however: setting $M$ excessively high (relative to $c$) may push critical values too far apart, reducing the reliability of monitoring decisions.

Practically, the PRS is calibrated so that shifts statistically equivalent to $n\delta$ cases per risk category signal a transition from acceptable to partial discrepancy. Similarly, shifts statistically equivalent to $nM\delta$ cases per risk category mark the shift from partial to full discrepancy. While $n\delta$ and 
$nM\delta$ provide easy reference points, actual deviations vary across categories --- some shifts less, others more --- necessitating a statistical procedure to assess overall resemblance. 

{A universal alternative is to compute $\delta$ using a uniform reference, i.e., setting $\bm{p}_0 = (1/B, \ldots, 1/B)$ solely for the purpose of defining $\delta$, while still using the actual $\bm{p}_0$ when evaluating the PRS statistic. This approach can be useful when a fixed benchmark is desired across portfolios, but may distort the interpretation of shift sizes, especially when some $p_{0,j}$ are small. In such cases, common in portfolios with small sample sizes or high concentration risk, the uniform threshold can misrepresent meaningful deviations.}

{The sensitivity parameters $\alpha_1$ and $\alpha_2$ should be chosen based on the organization's desired balance between under- and overreaction, ensuring consistent control over decision boundaries across monitoring regions: lower values prioritize model continuity, while higher values lead to increased sensitivity to population shifts. {Practitioners should caution against choosing excessively large values for $\alpha_1$ and $\alpha_2$ which may cause the intermediate monitoring region $\mathcal{R}_2$ to shrink or disappear entirely if the decision boundaries overlap. {Our findings suggest that values in the range $[0.01,0.2]$ generally yield stable results, though a trade-off may be necessary if {greater sensitivity} is required in one direction or the other. Furthermore, setting excessively large $ M $ can distort the framework by shifting the {critical values} too far apart, undermining the reliability of monitoring decisions.} Additionally, while $M$ and $c$ can be tailored to each portfolio being monitored, a more robust approach --- one that mitigates the risk of cherry-picking results --- is to fix these parameters across all portfolios, as $\delta$ already accounts for sample size considerations. This approach simplifies implementation while maintaining a systematic and objective monitoring framework.}

The PRS framework can be efficiently implemented using standard statistical software. Key computational steps include calculating $\delta$, determining $\lambda_{\mathrm{sup}}$ based on $B$, $n$, and $\delta$, and computing the decision-making {critical values} using non-central $\chi^2$ quantiles. The PRS is then compared to these {critical values} to assess model performance. {The diagram in Figure \ref{fig:flow_diag} illustrates the implementation flow of the PRS.} 
\vspace{0.5cm}

\begin{center}		\includegraphics[width=0.85\textwidth]{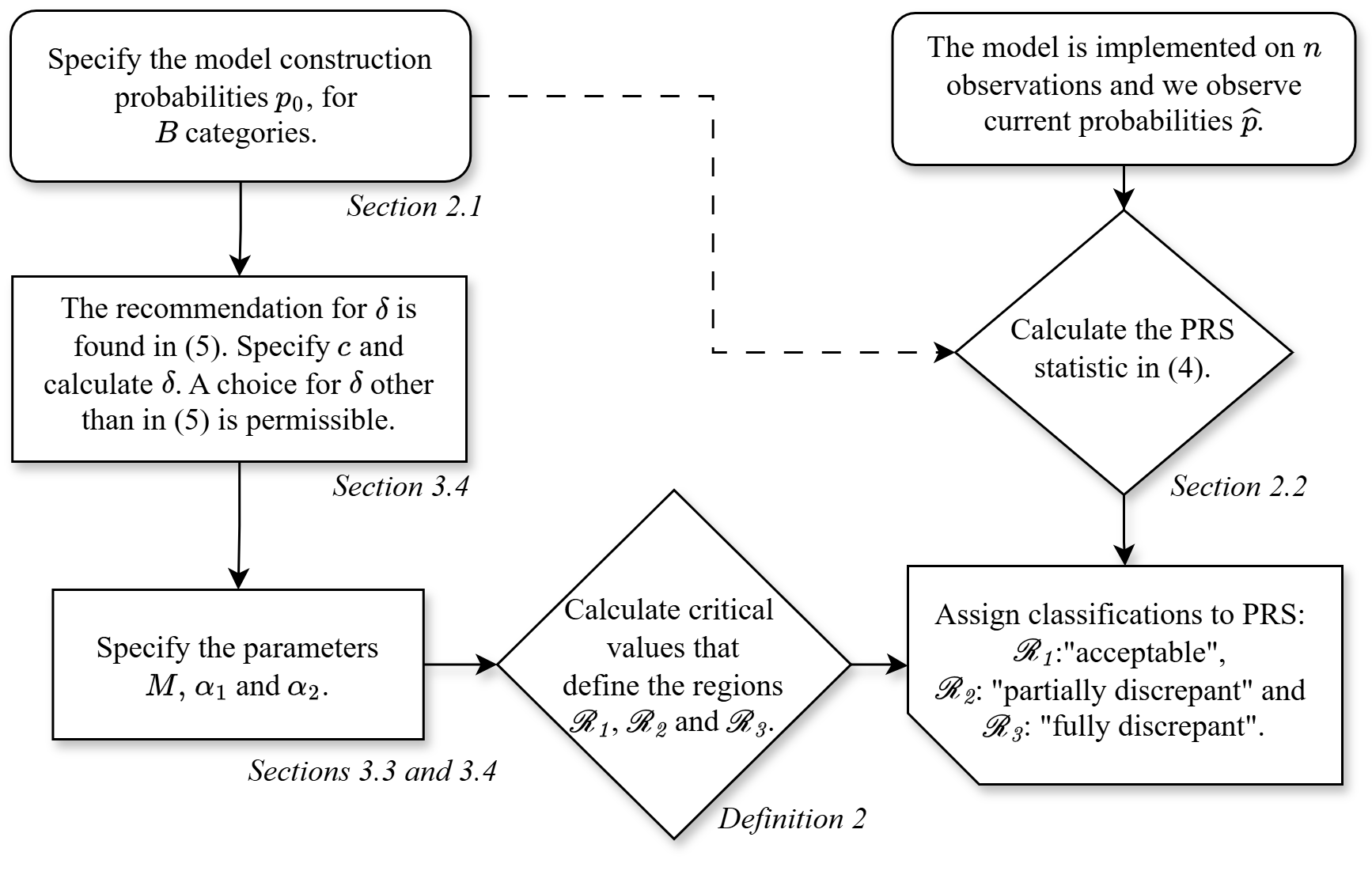} 
		\captionof{figure}{Flowchart depicting the implementation of the PRS method.}
        \label{fig:flow_diag}
\end{center}

\section{PRS method competency: A simulation study}  
\label{sec:PRS simulation}

To evaluate the performance of the PRS procedure, we conducted a comprehensive simulation study using a Monte Carlo approach. This study assesses the ability of the PRS method to distinguish between acceptable and discrepant deviations with respect to an equi-probable reference distribution under various sample sizes $ n $, numbers of categories $ B $, and specified degrees of deviation, denoted $ \delta_v $. 

For each scenario, the reference probability distribution was defined as $ \bm{p}_0 = (1/B, \dots, 1/B) $, representing $ B $ equally likely categories. A deviation of magnitude $ \delta_v $ was introduced to define perturbed probability distributions $ \bm{p}_v $. {Specifically, the first half of the categories have reduced probabilities $p_{v,j} = 1/B - \delta_v$ for $j = 1, \ldots, \lfloor B/2 \rfloor$, while the remaining categories have increased probabilities $p_{v,j} = 1/B + \delta_v$ for $j = \lceil B/2 \rceil + 1, \ldots, B$.} When $ B $ is odd, the central category remains unchanged at $ 1/B $. Under this setting, the PRS tolerance level defined in \eqref{eqn:rec_delta} becomes $\delta = c B^{-1} \sqrt{(B-1)/n}$ with specified scaling constant $c$. The boundary for unacceptable deviation is set at $ M\delta $ for specified $ M > 1 $.  Using the decision-making framework of Definition 2, we establish the PRS {critical values} $(\tau_1, \tau_2$), delineating the decision regions $(\mathcal{R}_1, \mathcal{R}_2, \mathcal{R}_3)$. 

We evaluated the PRS classification probabilities across 30 values of $ \delta_v $, ranging from no deviation, $\delta_v = 0$, to an extreme deviation, $ \delta_v = (3M+2) \delta $. This range captures shifts from within the acceptable range, where the population remains resemblant to the model, to larger deviations that are fully discrepant.

For each configuration of $ (n, B, \delta_v) $, we simulated $ K = 10^5 $ independent multinomial samples,  $\bm{X}_k \sim \text{Multinomial}(n, \bm{p}_v), \quad k = 1, \dots, K.$ For each sample, we estimated the empirical probability distribution $ \hat{\bm{p}}_k $ and computed the PRS statistic using $ \bm{p}_0 $ as the reference. The proportion of simulations falling into each of the three PRS decision regions was recorded. 

Simulations were conducted across a range of values for $ (n,B) $, varying $ 0.5 \leq c \leq 1 $, $ 1.2 < M < 2 $, and $ 0.01 \leq \alpha_1, \alpha_2 \leq 0.2 $. For illustrative purposes, results are presented for $(n,B) = (50,5)$ under two scenarios, firstly $(c,M)=(0.7,2)$ and $(\alpha_1,\alpha_2)=(0.05,0.1)$ --- see Figure \ref{fig:class_probs_n50_v1} --- and, secondly, $(c,M)=(1,1.6)$ and $(\alpha_1,\alpha_2)=(0.1,0.2)$ --- see Figure \ref{fig:class_probs_n50_v2}. This is to illustrate the versatility of the method across parameter choices. Finally, for $(n,B)=(10{,}000,20)$ with, $ (c,M)=(0.7,2)$ and $(\alpha_1,\alpha_2)=(0.05,0.1)$, we show the results in Figure \ref{fig:class_probs_n10000}. In each case, empirical classification probabilities are plotted against $\delta_v$ to visualize how classification behavior evolves across the deviation spectrum.

\begin{figure}[h!]
	\centering
		\centering
            \includegraphics[width=0.7\linewidth]{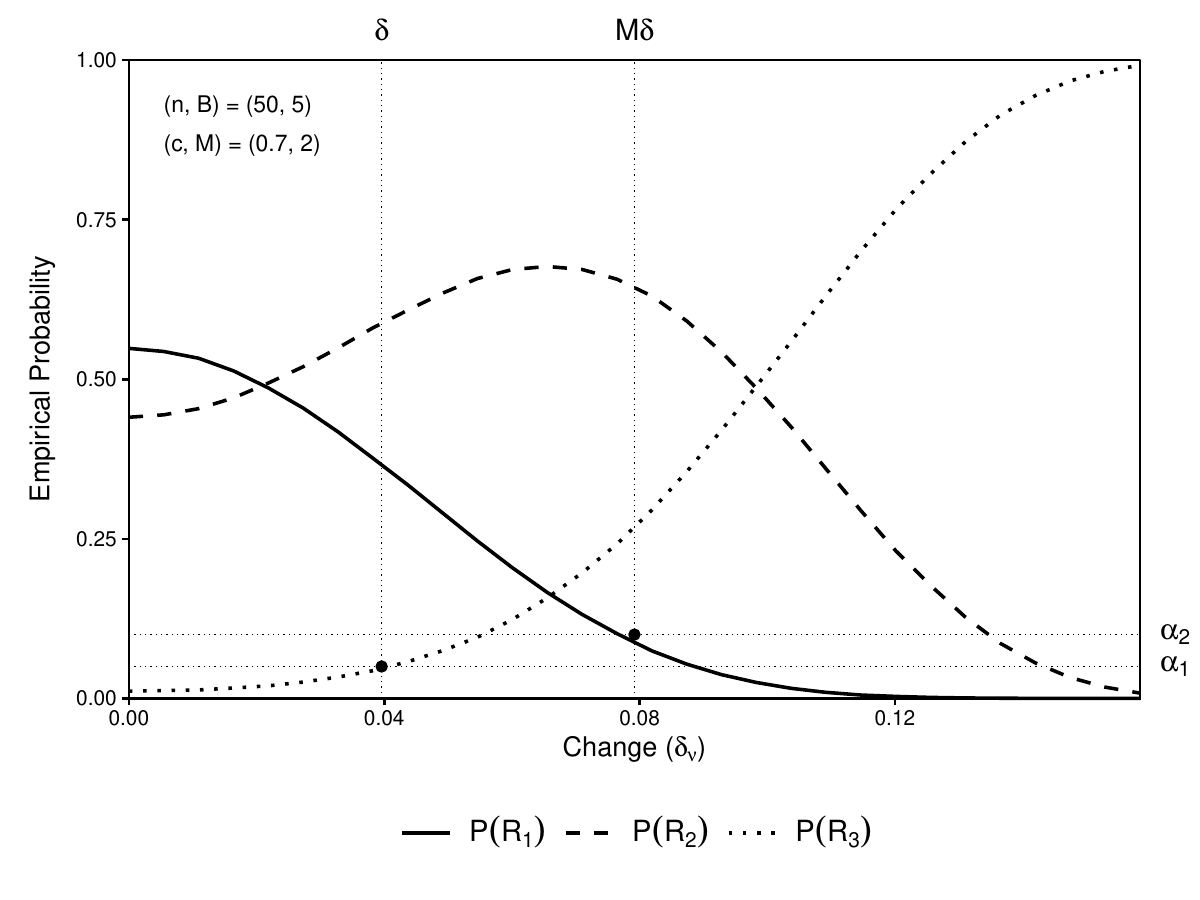}
    \caption{Empirical probabilities for $(n,B)=(50,5)$ with  $(\alpha_1,\alpha_2)=(0.05,0.1)$.}
    \label{fig:class_probs_n50_v1}
\end{figure}

\begin{figure}[h!]
		\centering
            \includegraphics[width=0.7\linewidth]{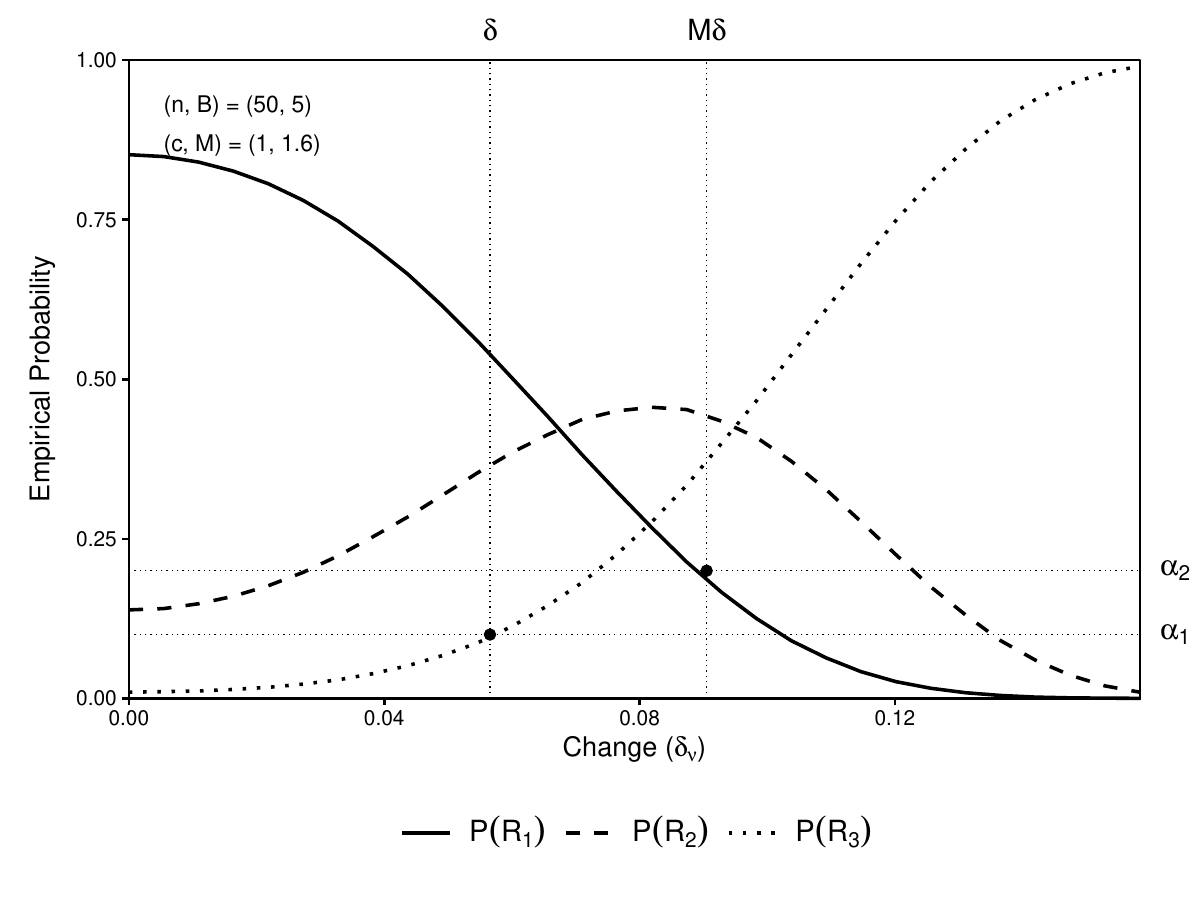}
    \caption{Empirical probabilities for $(n,B)=(50,5)$ with $(\alpha_1,\alpha_2)=(0.1,0.2)$.}
    \label{fig:class_probs_n50_v2}

\end{figure}

\begin{figure}
    \centering
    \includegraphics[width=0.7\linewidth]{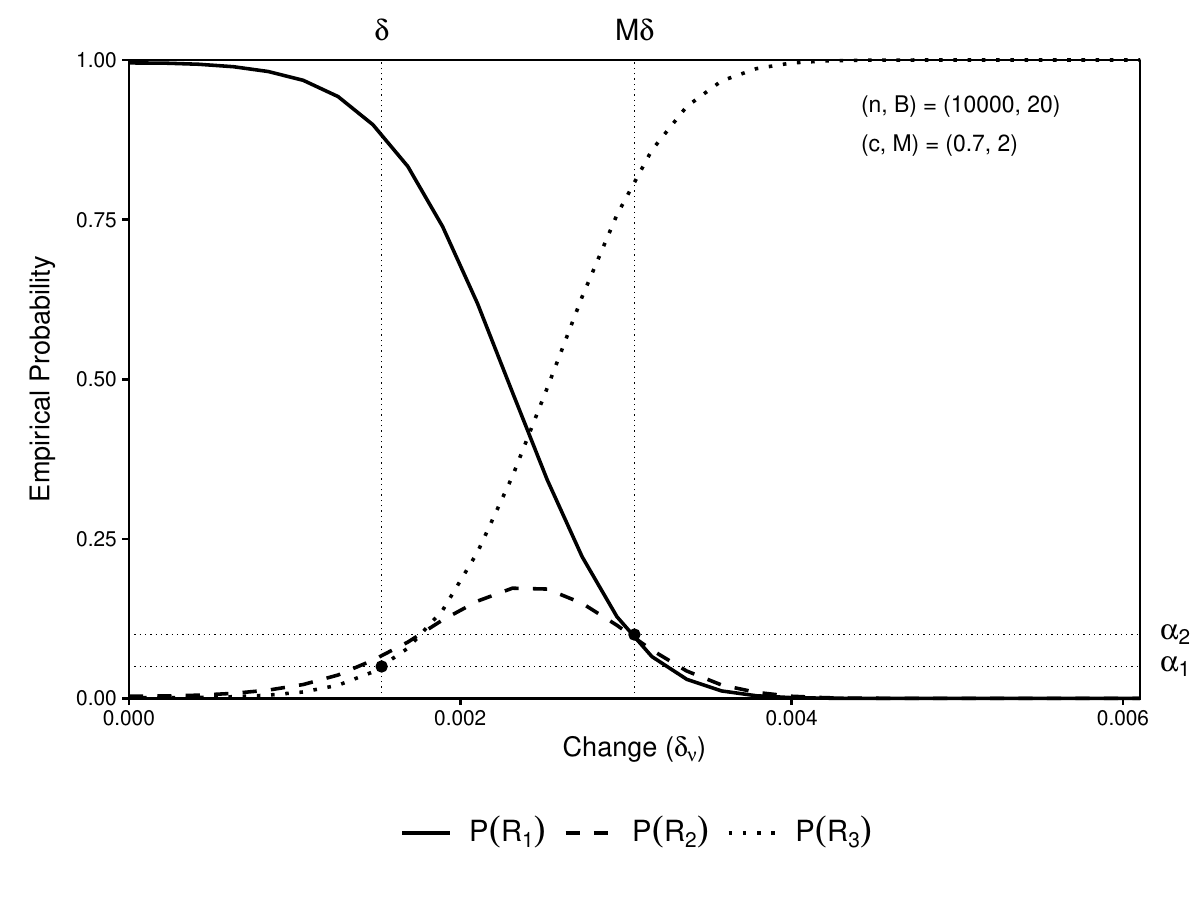}
    \caption{Empirical probabilities for $(n,B)=(10{,}000,20)$ with  $(\alpha_1,\alpha_2)=(0.05,0.1)$.}
    \label{fig:class_probs_n10000}
\end{figure}

\vspace{-0.5cm}
The results demonstrate that the PRS method effectively identifies when deviations become severe enough to warrant model reconstruction. When $ \delta_v = \delta $, the probability of PRS falling above the upper {critical value} $ \tau_2 $, $P(\mathcal{R}_3)$, closely matches the expected value of {$ \alpha_1$}. Similarly, under a deviation of size $ \delta_v = M\delta $, the probability of falling below $ \tau_1 $, $P(\mathcal{R}_1)$, remains near {$\alpha_2$}. These findings support the validity of the PRS framework and its reliability in classifying structured deviations from a multinomial reference model.

\FloatBarrier

\section{Banking applications and performance analysis}
\label{Application}

This section presents an empirical validation of the PRS based on a set of anonymized credit risk models using data from a large South African financial service provider. The dataset, spanning retail and corporate portfolios, was  selected to evaluate the PRS methodology across operationally relevant scenarios. Portfolio sizes have been rounded as part of the data anonymization. The examples encompass multiple portfolio configurations varying in sample size ($n$) and number of risk categories ($B$) where using equi-probable risk categories, ${p_{0i}}=1/B, \ i=1,...,B$, was deemed appropriate. Results are reported in Tables \ref{tab:n=2000 example} and \ref{tab:n=10000 B=20 example}. A red-amber-green ($\mathbbm{RAG}$) status was assigned using the decision regions in Definition~\ref{DefRegions}: red indicates full discrepancy, amber partial discrepancy, and green no significant shift. The designs considered are $(n,B)\in\{ (50,5),\, (500,10),\, (2{,}000,10),\, (10{,}000,20)\}$. 

{Note, category counts $n_i$ are tabulated rather than proportions $\hat{p}_i$ to enhance readability. This does not imply that $n$ remains fixed over time, nor that counts are the key statistic; the monitoring objective remains the evolution of $\hat{\bm{p}}$ relative to $\bm{p}_0$.}

We follow the procedure of Figure \ref{fig:flow_diag} to implement the PRS, using the recommended $\delta$ from \eqref{eqn:rec_delta} and choosing sensitivity parameters $\alpha_1=0.05$ and $\alpha_2=0.1$. These choices reflect risk aversion for an erroneous decision: $\alpha_1$ represents a $5\%$ probability of a red classification under $\delta$-resemblance, while $\alpha_2$ represents a $90\%$ probability of an amber or red classification under $M\delta$-resemblance. We used $c=0.7$ and $M=2$ for these applications. 

Conceptually, shifts statistically equivalent to $n\delta$ cases across all risk categories mark the transition from {green to amber}, while shifts statistically equivalent to $nM\delta$ cases across all risk categories mark the transition to {red}. For $(n,B) = (50,5)$ and $(n,B) = (10{,}000,20)$, these values are $1.98 \approx 2$ and $15.26 \approx 15$ cases per category for the transition to {amber}, and $3.96 \approx 4$ and $30.51 \approx 31$ cases per category {to red}. The critical values $\tau_1$ and $\tau_2$ delineating the regions {green, amber and red} are tabulated in Table \ref{tab:Critical values_PRS} along with the corresponding values of $\delta$. In all cases, $\delta$ (and $M\delta$) is smaller than $1/B$ ($= \min\limits_{j=1,\ldots,B} p_{0j}$), which is practically sensible.

We compare the PRS results with the widely used PSI, employing both the Lewis constants and the critical values proposed by \textcite{yurdakul2020statistical} (denoted by ``L'' and ``YN'', respectively). {We also include the discrete Kolmogorov-Smirnov (KS) test for further comparison.} {Of course, a direct comparison of the PRS with the PSI(YN) or the discrete KS is not entirely equivalent --- these are governed by the specification of two type I errors under a point null hypothesis of exact equality, in contrast to the nested composite structure of our proposed procedure. }Nonetheless, we choose upper and lower significance levels of $1\%$ and $10\%$: $p<1\%$ yields red, $p>10\%$ green, and intermediate values amber.}

\citereset
Moreover, it is known that PSI(YN) has low power at small sample sizes \parencite[Table~4]{yurdakul2020statistical}, possibly due to instability of the normalized PSI, as also seen in Figures~\ref{fig:mean_stab_ratios} and \ref{fig:var_stab_ratios}.  The YN-normed critical values are $\tau=2n^{-1}F^{-1}_{B-1}(1-\alpha)$ (in the notation of \textcite{yurdakul2020statistical} and assuming $m=n$), where $F_{\nu}^{-1}$ denotes the chi-square quantile function with $\nu$ degrees of freedom. 

Given the discrete nature of the KS statistic, particularly in small samples, it is not feasible to match critical values at specified significance levels and we rather report the p-value, $p(\mathrm{KS})$ from the \texttt{dgof}\footnote{The syntax used is \texttt{ks.test(X, ecdf(1:K), exact=F, simulate.p.value=T, B=10000)} where \texttt{X} is a numeric vector containing the repeated category number based on $\hat{p}$ and $B$ therein is the number of simulations and \texttt{K} the number of categories.} package in \texttt{R}, version 1.5.1 (2024/10/09), based on \textcite{RDGOF}. 

\begin{table}[h!]
  \centering
  \caption{ Critical values for the PRS, specifying $(c,M)=(0.7,2)$ and $(\alpha_1,\alpha_2)=(5\%,10\%)$, for $\delta$ from (\ref{eqn:rec_delta}). }
 \begin{tabular}{p{2cm}p{1.2cm}p{1.2cm}p{1.2cm}}
\toprule
\multicolumn{1}{c}{$(n,B)$} & \multicolumn{1}{c}{$\delta$} & \multicolumn{1}{c}{$\tau_1$} & \multicolumn{1}{c}{$\tau_2$}  \\
\cmidrule(lr){1-4}
$(50,5)$&0.039598&0.07441&0.25722\\
$(500,10)$&0.009391&0.03063&0.04890\\
$(2\ 000,10)$&0.004696&0.00766&0.01222\\
$(10\ 000,20)$&0.001526&0.00394&0.00439\\
\cmidrule(lr){1-4}
\end{tabular}
  \label{tab:Critical values_PRS}
\end{table}

\begin{table}[htbp]
  \centering
  \caption{Population resemblance comparison using the current population of size $n$ and observed category sample sizes $n_i$, $i=1,...,B$. }
     \begin{tabular}{p{0.2cm}p{6.3cm}p{1.7cm}p{1.7cm}p{1.7cm}}
\toprule
     3A & \multicolumn{1}{c}{${\mathbf{n_i}}$ with $n_{0i}=10$, $(n,B)=(50,5)$} & \multicolumn{1}{c}{PSI (L,YN)} & \multicolumn{1}{c}{PRS} & \multicolumn{1}{c}{p(KS)} \\
    \cmidrule(lr){1-5}
    $t_1$ & (6, 9, 10, 11, 14)& 0.072 ($\mathbbm{G,G}$) & 0.068 ($\mathbbm{G}$)& 0.401 ($\mathbbm{G}$)\\
    $t_2$ & (4, 10, 11, 11, 14)& 0.141  ($\mathbbm{A,G}$) & 0.108 ($\mathbbm{A}$)& 0.232 ($\mathbbm{G}$)\\
    $t_3$ & (7, 8, 8, 10, 17)& 0.114 ($\mathbbm{A,G}$) &  0.132 ($\mathbbm{A}$)& 0.125 ($\mathbbm{G}$)\\
    $t_4$ & (3, 8, 12, 13, 14)& 0.227  ($\mathbbm{A,G}$) & 0.164 ($\mathbbm{A}$)& 0.027 ($\mathbbm{A}$)\\
    $t_5$ & (2, 9, 12, 13, 14)& 0.310  ($\mathbbm{R,G}$) & 0.188 ($\mathbbm{A}$)& 0.028 ($\mathbbm{A}$)\\
    $t_6$ & (2, 5, 13, 14, 16)& 0.426  ($\mathbbm{R,A}$) & 0.300 ($\mathbbm{R}$)& $<0.001$ ($\mathbbm{R}$)\\
 
      \cmidrule(lr){1-5}

     3B & \multicolumn{1}{c}{${\mathbf{n_i}}$ with $n_{0i}=50$, $(n,B)=(500,10)$} & \multicolumn{1}{c}{PSI (L,YN)} & \multicolumn{1}{c}{PRS} & \multicolumn{1}{c}{p(KS)} \\
    \cmidrule(lr){1-5}

    $t_1$ & \Centerstack{(35, 40, 45, 45, 47, 50, 55, 58, 60, 65)} & 0.032 ($\mathbbm{G,G}$) & 0.032 ($\mathbbm{A}$)& 0.002 ($\mathbbm{R}$)\\
    $t_2$ & \Centerstack{(40, 45, 45, 45, 47, 48, 55, 55, 60, 60)} & 0.017 ($\mathbbm{G,G}$) & 0.018 ($\mathbbm{G}$)& 0.024 ($\mathbbm{A}$)\\
    $t_3$ & \Centerstack{(35, 36, 42, 43, 44, 44, 60, 60, 61, 75)} & 0.060 ($\mathbbm{G,A}$) & 0.062 ($\mathbbm{R}$)& $<0.001$ ($\mathbbm{R}$)\\
    $t_4$ & \Centerstack{(20, 35, 35, 40, 40, 62, 65, 65, 65, 73)} & 0.131 ($\mathbbm{A,R}$)& 0.116 ($\mathbbm{R}$)& $<0.001$ ($\mathbbm{R}$)\\
    \cmidrule(lr){1-5}
     3C & \multicolumn{1}{c}{${\mathbf{n_i}}$ with $n_{0i}=200$, $(n,B)=(2{,}000,10)$} & \multicolumn{1}{c}{PSI (L,YN)} & \multicolumn{1}{c}{PRS} & \multicolumn{1}{c}{p(KS)} \\
    \cmidrule(lr){1-5}

    $t_1$ & \Centerstack{(160, 170, 180, 180, 190, 200, 210, 220, 240, 250)} & 0.020 ($\mathbbm{G,A}$) & 0.020 ($\mathbbm{R}$)& $<0.001$ ($\mathbbm{R}$)\\
    $t_2$ & \Centerstack{(180, 180, 184, 190, 194, 200, 200, 210, 222, 240)} & 0.008 ($\mathbbm{G,G}$) & 0.008 ($\mathbbm{A}$)& 0.004 ($\mathbbm{R}$)\\
    $t_3$ & \Centerstack{(180, 180, 190, 194, 200, 200, 204, 210, 220, 222)} & 0.005 ($\mathbbm{G,G}$) & 0.005 ($\mathbbm{G}$)& 0.035 ($\mathbbm{A}$)\\
    $t_4$ & \Centerstack{(160, 170, 170, 178, 180, 210, 210, 220, 242, 260)} & 0.025 ($\mathbbm{G,R}$) & 0.026 ($\mathbbm{R}$)& $<0.001$ ($\mathbbm{R}$)\\

    \cmidrule(lr){1-5}
    \end{tabular}
  \label{tab:n=2000 example}
\end{table}

\begin{table}[h!]
  \centering
  \caption{Population resemblance comparison using the current population of size $n=10{,}000$ and observed category sample sizes $n_i$, $i=1,...,B$, $B=20$ and $n_{0i}=500$.}
     \begin{tabular}{p{0.2cm}p{7cm}p{1.8cm}p{1.5cm}}
\toprule
     & \multicolumn{1}{c}{${\mathbf{n_i}}$} & \multicolumn{1}{c}{PSI (L,YN)} & \multicolumn{1}{c}{PRS}  \\
    \cmidrule(lr){1-4}

    $t_1$ & 
    \Centerstack{(425, 455, 480, 480, 480, 480, 485, 491, 495, 495,\\500, 502, 502, 502, 502, 520, 540, 546, 550, 570)} & 0.0042 ($\mathbbm{G,G}$) & 0.0042 ($\mathbbm{A}$)\\ \hdashline
    $t_2$ & \Centerstack{(150, 170, 400, 400, 450, 450, 460, 460, 525, 525,\\545, 545, 550, 550, 600, 620, 650, 650, 650, 650)} & 0.1060 ($\mathbbm{A,R}$) & 0.0769 ($\mathbbm{R}$)\\ \hdashline
    $t_3$ & \Centerstack{(445, 455, 480, 480, 485, 485, 490, 495, 500, 500,\\501, 502, 502, 510, 510, 520, 520, 530, 540, 550)} & 0.0025 ($\mathbbm{G,G}$) & 0.0025 ($\mathbbm{G}$)\\ \hdashline
    $t_4$ & \Centerstack{(425, 425, 440, 440, 445, 445, 460, 460, 475, 475,\\490, 490, 525, 525, 555, 555, 585, 585, 600, 600)} & 0.0139 ($\mathbbm{G,R}$) & 0.0142 ($\mathbbm{R}$)\\ \hdashline
    $t_5$ & \Centerstack{(390, 390, 450, 450, 450, 450, 460, 460, 475, 475,\\525, 525, 545, 545, 550, 550, 555, 555, 600, 600)} & 0.0153 ($\mathbbm{G,R}$) & 0.0150 ($\mathbbm{R}$)\\ \hdashline
    $t_6$ & \Centerstack{(440, 465, 465, 475, 475, 480, 480, 485, 485, 488,\\490, 490, 510, 510, 520, 520, 550, 550, 550, 572)} & 0.0045 ($\mathbbm{G,G}$) & 0.0045 ($\mathbbm{R}$)\\
       \cmidrule(lr){1-4}
        \multicolumn{4}{l}{Note: In all of $t_1,\dots,t_6$, $p(KS)<0.001$ ($\mathbbm{R}$).}\\
    \cmidrule(lr){1-4}
    \end{tabular}  \label{tab:n=10000 B=20 example}
\end{table}

When comparing the PSI(L) with the PRS, the results are commensurate with the conclusions from Table \ref{TABLE1}: the PRS procedure less frequently indicates ``\textit{full discrepancy}'' (red) in small samples ($n=50$), while doing so more frequently in larger samples ($n\geq 500$). The same conclusion holds when comparing the PSI(L) with the PSI(YN). Recall that the PSI(L) has an inflated {probability of indicating a shift when none has occurred} in small samples and, conversely, a too-close-to-zero {probability of indicating ``\textit{full discrepancy}'' (red) in large samples. Notably, in the cases where $n$ exceeds $500$, PSI(L) seems insensitive to both partial and full \textit{discrepancy} (amber and red). Observe in all cases considered here, that the PRS indicates \textit{discrepancy}  (amber and red) more frequently than the PSI(YN). 

If we considered an alternative setup for the methodology, say $c=0.9$, $M=1.5$ and $\alpha_1=0.1$, $\alpha_2=0.2$, the only differently assigned statuses (now green i.s.o. amber) would occur at $t_1$ of $(500,10)$, $t_2$ of $(n,B)=(2{,}000,10)$; as well as an amber status instead of red at $t_6$ of $(10{,}000,20)$. Here, the amber region is smaller, as evident from the choices of $M$ and larger values of $\alpha_1$ and $\alpha_2$. Nonetheless, there is a large degree of overlap between the two sets of results. 

A comparison with the discrete KS remains. As with the PSI(YN), the KS is based on a null hypothesis of exact equality, while the PRS includes a risk tolerance through $\delta$. The KS is also not distribution-free with respect to $\bm{p}_0$, a further drawback. As a first departure, observe Table \ref{tab:n=2000 example}A: the PRS is more likely to indicate \textit{discrepancy} (red or amber) than the KS. Further to this, it is well known that the power of the KS is small for small sample sizes (not unlike the PSI using YN critical values) possibly explaining the more frequent green status where $n=50$, compared to the larger sample sizes. In all other cases (Tables \ref{tab:n=2000 example}B, \ref{tab:n=2000 example}C and \ref{tab:n=10000 B=20 example}), the KS signals full discrepancy (red) almost always. This is a direct result of the substantial power of the KS at larger sample sizes, and by its design, that any shift away from the null will swiftly be detected. Perhaps, detecting {even the slightest} shifts so often might not be ideal to a risk practitioner. Utilizing the PRS in these cases allows for a range of detection capabilities (see case $t_1$ and $t_3$ of $(n,B)=(10{,}000,20)$ with an amber and green status, respectively).

We conclude from this comparative real-world study that, among the measures considered here, {the PRS is universally competent at a range of sample sizes, including small sample sizes. A clear advantage of the PRS over both the PSI and KS is the inclusion of the concept of $\delta$-resemblance and the tuning parameters $c$ and $M$ allowing the practitioner to calibrate the procedure to align with their risk appetite. The PRS clearly indicates \textit{discrepancy} sufficiently in smaller samples and often enough in larger samples.} Unlike the KS, the PRS is sensitive to ranges of shifts over multiple risk categories and has easily obtainable critical values that are unique in small sample sizes.

\section{Conclusion}
\label{sec:Conclusion}

Monitoring for changes in the population underlying a developed model is a common practice, especially in credit risk modeling.  Over the years, several measures --- most notably the PSI --- have been proposed. However, limitations in these methods have spurred research into alternative approaches for assessing population resemblance. Our contribution, the PRS, utilizes the Pearson chi-square statistic and non-central chi-square distribution to address these gaps. 

The main advantageous features are that the PRS accommodates sample-size dependent critical values and its explicit specification of risk tolerance. The PRS is statistically well-founded, performing reliably across a range of sample sizes, including in small samples. Unlike the discrete Kolmogorov-Smirnov test, the PRS is asymptotically distribution-free with respect to $\bm{p}_0$. Further, the PRS critical values are designed to account for acceptable levels of population shift, aligning with a composite null hypothesis framework. The risk tolerance $\delta$ explicitly incorporates an assumption of limited population shift, distinguishing the PRS from the KS test and the PSI of Lewis and \textcite{yurdakul2020statistical}, which do not accommodate this structured flexibility. The tolerance parameter is incorporated through the concept of $\delta$-resemblance, a convenient way to communication population shift to practitioners in using the concept to derive business outcomes. 

These competency characteristics of the PRS were demonstrated through Monte Carlo simulations and real-world applications. In the applications, the suitability of the PRS was showcased, measuring the resemblance between populations given both small sample sizes (often encountered in low default portfolios) and in larger samples frequently encountered in retail portfolios of a bank. We have clearly shown that the PRS indicates (partial and/or full) discrepancy sufficiently in smaller samples and often enough in larger samples. 

Future research could explore relaxing the assumption of fixed reference probabilities $\bm{p}_0$ by adopting a two-sample framework where $\bm{p}_0$ arises through random sampling. Another practical direction would be redefining the tolerance parameter $\delta$ to account for varying risk category costs, leading to a multivariate formulation that reflects their relative importance.

{To conclude, we gratefully acknowledge the reviewers for their insightful and constructive feedback, which significantly strengthened the manuscript. Their comments helped clarify the composite null hypothesis framework, refine the exposition of the proposed procedure, suggest connections to stochastic ordering, and led to a more general and rigorous formulation and proof of Proposition 2.}\bigskip

\section{Compliance with ethical standards}

All authors certify that they have no affiliations with or involvement in any organization or entity with any financial interest or non-financial interest in the subject matter or materials discussed in this manuscript. No funds, grants, or other support was received.

\setstretch{1.46} 
\setstretch{1.667} 
\printbibliography

\end{document}